\documentclass[aps,final,notitlepage,oneside,twocolumn,nobibnotes,nofootinbib,
superscriptaddress,noshowpacs,centertags]{revtex4-1}

\usepackage[cp1251]{inputenc}
\usepackage[english]{babel}
\usepackage{graphicx}
\usepackage{latexsym}
\usepackage{amssymb}
\usepackage{amsmath}
\usepackage{float}

\begin{document}

\title{Event horizon image within black hole shadow}
\author{V. I. Dokuchaev}\thanks{e-mail: dokuchaev@inr.ac.ru}
\affiliation{Institute for Nuclear Research, Russian Academy of Sciences,
pr. 60-letiya Oktyabrya 7a, Moscow, 117312 Russia}
\affiliation{MEPhI National Research Nuclear University, Kashirskoe sh. 31, Moscow, 115409 Russia}
\author{N. O. Nazarova}\thanks{e-mail: natalia.nazarova@sissa.it}
\affiliation{MEPhI National Research Nuclear University, Kashirskoe sh. 31, Moscow, 115409 Russia}
\affiliation{Scuola Internazionale Superiore di Studi Avanzati (SISSA), Via Bonomea 265, 34136 Trieste (TS) Italy}

%\date{\today}

\begin{abstract}
We argue that a genuine image of the black hole viewed by a distant observer is not its shadow, but a more compact event horizon image probed by the luminous matter plunging into black hole. The external border of the black hole shadow is washed out by radiation from matter plunging into black hole and approaching the event horizon. This effect will crucially influence the results of future observations by the Event Horizon Telescope. We show that gravitational lensing of the luminous matter plunging into black hole provides the possibility for visualization of the event horizon. The lensed image of the event horizon is formed by the highly red-shifted photons emitted by the plunging matter very near the black hole event horizon and detected by a distant observer. The resulting event horizon image is a gravitationally lensed projection on the celestial sphere of the whole black hole event horizon sphere. Seemingly, black holes are the unique objects in the Universe which may be viewed by distant observers at once from both the front and back sides.
\end{abstract}

\maketitle 

\section{Introduction}

The supermassive black hole SgrA* at the Galactic Center with a mass $M=(4.3\pm0.3)10^6M_\odot$ \cite{Gillessen09,Meyer12,Johnson15,Chatzopoulos15,Johannsen16c,Eckart17} is the object of especially steadfast investigations. The one reason is that this black hole is the nearest ``dormant'' quasar dwelling precisely at the center of our native Galaxy. The other reason is that nowadays there are technologies permitting to view the enigmatic black hole directly for the first time. The Event Horizon Telescope, a global network of mm and sub-mm telescopes \cite{Fish16,Lacroix13,Kamruddin,Johannsen16,Johannsen16b,Broderick16,Chael16,Kim16,Roelofs17,Doeleman17},
intends to reveal the shadow of supermassive black hole at the Galactic Center \cite{Synge,49,Chandra,51,52,Falcke13,57,Cunha15,Abdujabbarov15,Younsi16,70,Thorne2015}. It would be the first direct experimental verification (or falsification) of the very existence of black holes in the Universe. 

The black hole shadow is formed by gravitational lensing of luminous background, consisting of bright stars or hot gas, which are located either far enough from the black hole or at the stationary orbits around it. In other words, the black hole shadow is a capture cross section of background photons viewed by distant observers on the celestial sphere. For more general and rigorous definitions of the black hole shadow see, e.\,g., \cite{Grenzebach14,Grenzebach15,Cunha18} and references therein.

A next stage in investigation of the black hole shadow at the Galactic Center would be the detailed elaboration of its shape for verification of the General Relativity in the strong field limit  \cite{Grenzebach14,Grenzebach15,Cunha18,71,65,Dokuch14,DokEr15,FizLab,Nucamendi15,Nucamendi16,CliffWill17a,CliffWill17b,Goddi17}. One of the promising interferometric projects for this future study is the proposed Russian Millimetron space interferometer project with an angular resolution of about one nanoarcsecond \cite{64}. Realization of the Millimetron space interferometer project will provide the basis for realization of even the more advanced projects for verification not only the General Relativity in the strong field limit but also its numerous modifications.

In view of the very weak accretion onto the supermassive black hole SgrA* at the Galactic Center we will neglect in the following the difference between the apparent and event horizons. It is generally believed that the black hole event horizon is invisible and cannot be imaged. We show below that gravitational lensing of the luminous matter plunging into black hole provides the principle possibility for revealing the event horizon image. This event horizon image is projected on the celestial sphere within the position of black hole shadow, predicted by the Kerr metric. The resulting event horizon image is a gravitationally lensed projection on the celestial sphere of the whole black hole event horizon sphere. 

The black hole shadow and the event horizon image may be revealed at once by simultaneous observations with several telescopes of different types. Namely, the black hole shadow may be detected in the X-rays or in the near far Infra-Red by observation of the foreground sources and the gravitationally lensed images, e.\,g., of normal stars or neutron stars behind the black hole, which are moving around black hole on the stationary orbits. At the same time, the distant observer may reveal the event horizon image by detecting the highly red-shifted photons in the radio waves, radiated very near the black hole horizon by the plunging compact sources, e.\,g., neutron stars or gas clouds.

It is convenient to use units with the gravitational constant $G=1$ and the velocity of light $c=1$. With these units we define the dimensionless values of space distances, $r\Rightarrow r/M$, and time intervals, $t\Rightarrow t/M$ and the dimensionless spin parameter of the Kerr space-time for rotating black hole, $0\leq a=J/M^2\leq1$, where $M$ and $J$ are, respectively, the black hole mass and angular momentum. The event horizon radius in the Kerr space-time is $r_{\rm h}=1+\sqrt{1-a^2}$. At $a=0$ the Kerr space-time coincides with the Schwarzschild one. For numerical calculations of gravitational lensing in the Kerr space-time background we use the integral equations of motion for photons \cite{Chandra,Carter68,deFelice,BPT,MTW} in Boyer--Lindquist coordinates \cite{BoyerLindquist}: 
\begin{equation}\label{eq2425}
\int\limits_C\frac{dr}{\sqrt{V_r}}=\int\limits_C\frac{d\theta}{\sqrt{V_\theta}},
\end{equation}
\begin{equation}
\phi=\!\!\int\limits_C\frac{a[E(r^2\!+\!a^2)\!-\!La]}{(r^2-2r+a^2)\sqrt{V_r}}\,dr
+\!\!\int\limits_C\frac{L\!-\!aE\sin^2\theta}{\sin^2\theta\sqrt{V_\theta}}\,d\theta,
\label{eq25b} %\nonumber
\end{equation}
where
\begin{eqnarray}
V_r &=& [E(r^2\!+\!a^2)\!-\!a L]^2-\Delta[\mu^2r^2+(L\!-\!aE)^2+Q], \label{Vr} \\
V_\theta &=& Q-\cos^2\theta[a^2(\mu^2-E^2)+L^2\sin^{-2}\theta]
\label{Vtheta} 
\end{eqnarray}
are, respectively, the latitudinal and radial effective potentials. There are four constants of motion for a test particle in the Kerr metric: mass $\mu$, total energy $E$, azimuthal angular momentum $L$ and Carter constant $Q$, defining the non equatorial motion. The integrals in (\ref{eq2425}) and (\ref{eq25b}) are understood to be path integrals along the test particle trajectory connecting a source to an observer.

Trajectories of massive particles ($\mu\neq0$) in the Kerr space-time are determined by three parameters (constants of motion): $\gamma=E/\mu$, $\lambda=L/E$ and $q=\sqrt{Q}/E$. The corresponding trajectories of photons (null geodesics) are determined by two parameters, $\lambda$ and $q$, which coincide with the horizontal and vertical impact parameters of photons on the celestial plane for a distant observer in the black hole equatorial plane. The Boyer--Lindquist coordinates, which we use in numerical calculations of photon trajectories, coincide asymptotically at $r\to\infty$ with the flat Euclidean spherical coordinates. 

See \cite{Gralla15,Strom16,Gralla16,Strom17,Strom17b} for corresponding examples of analytical computations for null geodesics in the Kerr space-time. An example of numerical gravitational lensing animation of the compact star orbiting the rotating black hole see at \cite{doknaz18a,doknaz18b}. 

\begin{figure}%[h]
	\begin{center}
		\includegraphics[angle=0,width=0.45\textwidth]{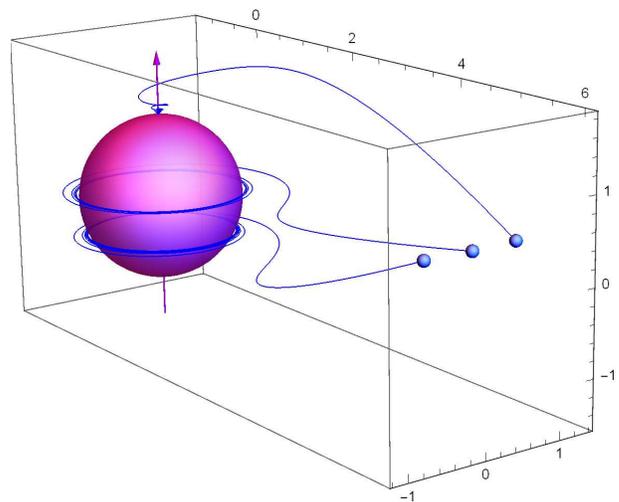}
	\end{center}
	\caption{Examples of $3D$ trajectories in the Boyer-Lindquist coordinates for massive test particles ($\mu\neq0$), plunging into the fast rotating back hole ($a=0.998$) near the north pole of the event horizon ($\gamma=1$, $\lambda=0$, $q=1.85$), near its equator ($\gamma=1$, $\lambda=-1.31$, $q=0.13$) and in the south hemisphere ($\gamma=1$, $\lambda=-1.31$, $q=0.97$}
	\label{fig1}
\end{figure}

See in Fig.~1 the examples of $3D$ trajectories for massive test particles, plunging into the fast rotating back hole and infinitely ``winding'' around black hole by approaching to the event horizon. These $3D$ trajectories are calculated from combined numerical solution of integral equations of motion (\ref{eq2425}) and (\ref{eq25b}).

\begin{figure}%[h]
	\begin{center}
		\includegraphics[angle=0,width=0.45\textwidth]{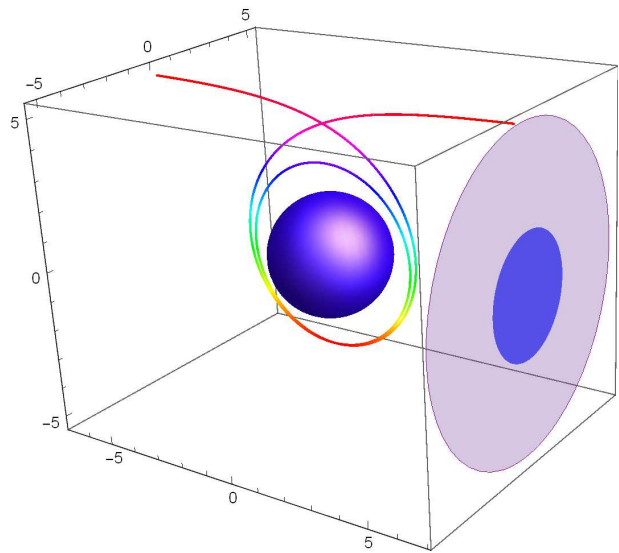}
	\end{center}
	\caption{The shadow (purple disk) of Schwarzschild black hole ($a=0$) with a radius $r_{\rm sh}=3\sqrt{3}\simeq5.196$. Inside the shadow it is shown the fictitious image (blue disk) of the event horizon with radius $r_h=2$ in the imaginary Euclidean space. It shown the photon trajectory (multicolored $3D$ curve) near the shadow boundary (with impact parameters $\lambda=0$ and $q=3\sqrt{3}$), starting from a distant background, winding near a radial turning point at $r_{\rm min}=3$ around the black hole event horizon (blue sphere) and reaching finally a distant observer at the north pole of the black hole shadow.}
	\label{fig2}
\end{figure}

\begin{figure}%[h]
	\begin{center}
		\includegraphics[angle=0,width=0.45\textwidth]{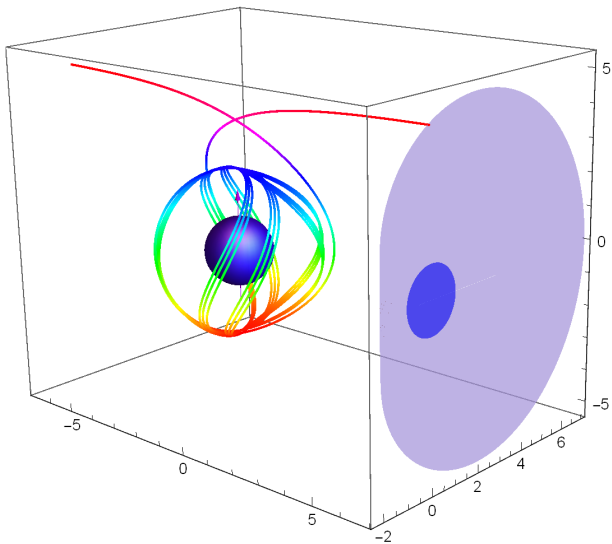}
	\end{center}
	\caption{The shadow (purple region) of the extremely fast rotating Kerr black hole ($a=1$). As a representative example it is shown the photon trajectory (multicolored $3D$ curve) near the shadow boundary (with $\lambda=0$ and  $q=\sqrt{3-\sqrt{2}}(1+\sqrt{2})^{3/2}]\simeq4.72$), starting from a distant background, winding around the black hole event horizon (blue sphere) and reaching a distant observer at the intersection of the shadow border with the projection of the black hole rotation axis. This photon has a radial turning point at $r_{\rm min}=1+\sqrt{2}$. A vertical magenta arrow is the black hole rotation axis. Inside the shadow it is shown the fictitious image (blue disk) of the event horizon with radius $r_h=1$ in the imaginary Euclidean space.}
	\label{fig3}
\end{figure}

\section{Black hole shadow}

The apparent shape of the black hole shadow, as seen by a distant observer in the equatorial plane, is determined parametrically, $(\lambda,q)=(\lambda(r),q(r))$, from simultaneous solution of equations $V_r(r)=0$ and $[rV_r(r)]'=0$ (see, e.\,g.,\cite{49,Chandra}):
\begin{eqnarray} \label{shadow1}
\lambda&=&\frac{-r^3+3r^2-a^2(r+1)}{a(r-1)}, \\
q^2&=&\frac{r^3[4a^2-r(r-3)^2]}{a^2(r-1)^2}.
\label{shadow2}
\end{eqnarray}

The shadow of Schwarzschild black hole $(a=0)$ is shown in Fig.~2. The largest (purple) disk is the black hole shadow with a radius $r_{\rm sh}=3\sqrt{3}\simeq5.196$. The blue disk at the shadow center is a fictitious event horizon image in the imaginary Euclidean space. 

The corresponding shadow of extreme Kerr black hole $(a=1)$ is shown in Fig.~3. It is also shown the photon trajectory (multicolored $3D$ curve) near the shadow boundary (with $\lambda=0$ and $q=\sqrt{3-\sqrt{2}}(1+\sqrt{2})^{3/2}]\simeq4.72$), starting from a distant background and reaching distant observer at the intersection of the shadow border with the projection of the black hole rotation axis. This photon has a radial turning point at $r_{\rm min}=1+\sqrt{2}$. Inside the shadow it is shown the fictitious image (blue disk) of the event horizon with radius $r_h=1$ in the imaginary Euclidean space.

\section{Event horizon image}

\begin{figure}%[h]
	\begin{center}
		\includegraphics[angle=0,width=0.45\textwidth]{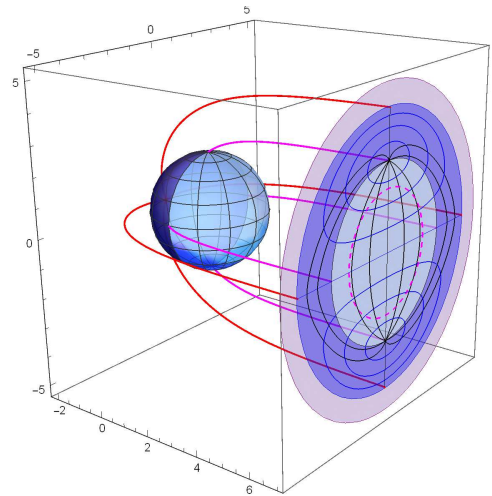}
	\end{center}
	\caption{The event horizon image (blue disk) inside the black hole shadow (purple disk) in the Schwarzschild case ($a=0$). The dashed magenta circle here and on the subsequent Figs. is a fictitious outer border of the event horizon image in the imaginary Euclidean space. The trajectories of some photons forming the event horizon image are shown: 4 (red colored) trajectories with $\sqrt{\lambda^2+q^2}=r_{\rm eh}$ are starting from the farthest point of the event horizon globe, and,respectively, 4 (magenta colored) trajectories with $\sqrt{\lambda^2+q^2}=r_{\rm EW}$ are starting from the East-West meridian. Images of some parallels (blue curves) and meridians (black curves) are also shown on the event horizon globe (blue sphere) and on its projected image (blue region) viewed by distant observer. The light-blue part of the event horizon image is a projection of the nearest event horizon hemisphere. Respectively, the dark-blue part of the event horizon image is a projection of the farthest event horizon hemisphere. A distant observer views the event horizon at once from both the front and back sides.}
	\label{fig4}
\end{figure} 
\begin{figure}%[h]
	\begin{center}
		\includegraphics[angle=0,width=0.45\textwidth]{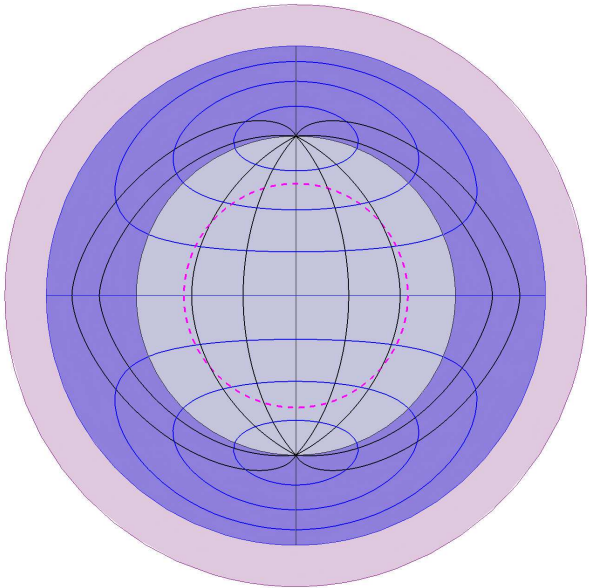}
	\end{center}
	\caption{The detailed view of the event horizon image inside black hole shadow in the Schwarzschild case ($a=0$).}
	\label{fig5}
\end{figure} 

\begin{figure}%[h]
	\begin{center}
		\includegraphics[angle=0,width=0.45\textwidth]{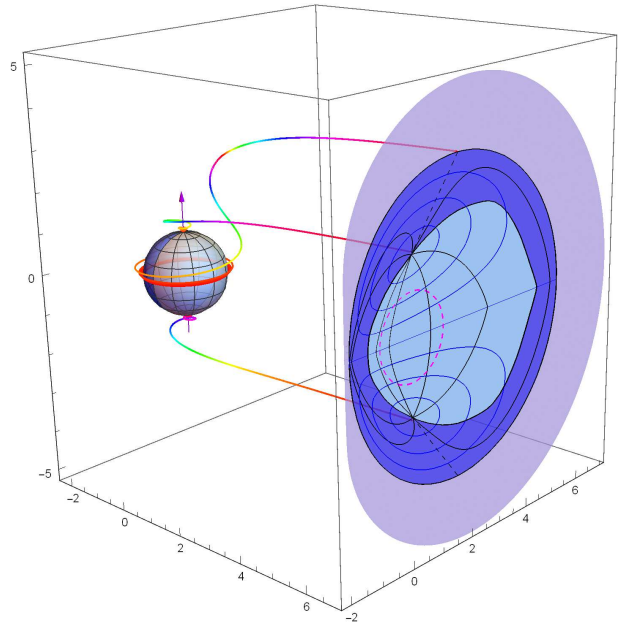}
	\end{center}
	\caption{The event horizon image of extreme Kerr black hole ($a=1$) and trajectories  (multicolored $3D$ curves) of some photons forming this image and outgoing to a distant observer from the North and South poles (with $\lambda=0$, $q = 1.77$) and from the equator (with $\lambda=-1.493$, $q=3.629$) of the event horizon globe with a radius $r_h=1$. The largest purple region is the black hole shadow. Images of some parallels (blue curves) and meridians (black curves) are also shown on the event horizon globe (blue sphere) and on its projected image (blue region) viewed by distant observer. The light-blue part of the event horizon image is a projection of the nearest event horizon hemisphere. Respectively, the dark-blue part of the event horizon image is a projection of the farthest event horizon hemisphere. A distant observer views the event horizon at once from both the front and back sides.}
	\label{fig6}
\end{figure} 
\begin{figure}%[h]
	\begin{center}
		\includegraphics[angle=0,width=0.45\textwidth]{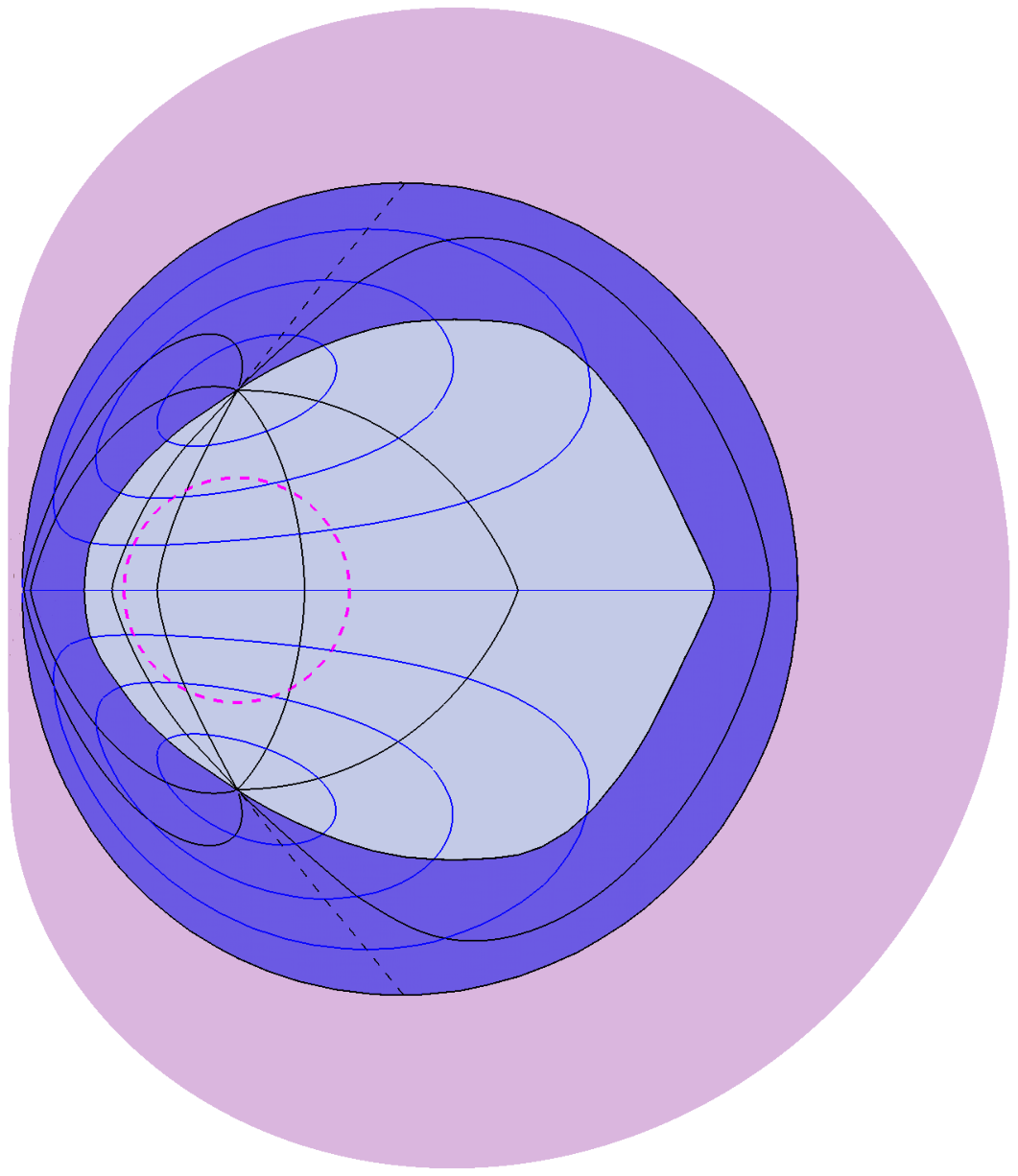}
	\end{center}
	\caption{The detailed view of the event horizon image of extreme Kerr black hole. in the extreme Kerr case ($a=1$).}
	\label{fig7}
\end{figure} 

In the vicinity of a black hole, like the SgrA*, besides the near stationary foreground matter, there may be the non-stationary luminous objects (dense gas clouds and compact stars), plunging into black hole. These luminous objects will inevitably washing out the outer border of the observable black hole shadow. 

It is useful to imagine the gedankenexperiment of throwing the luminous probes in all directions and from every quarter into the black hole. A distant observer will see the progressively weakened image of each plunging probe, approaching the event horizon with the redshift of emitting photons growing to infinity for the ``last'' photon. Position of the last lensed photon on the celestial sky, determined by the tracking of separate plunging probe, will map the single point on the event horizon. The set of these points would be the one-to-one projection of the whole event horizon sphere into its lensed image. The last detected highly red-shifted photons will mark the specific final points on the event horizon image. 

In result, the limiting process of registration the progressively fading images of plunging luminous probes (like the compact stars and hot clouds) provides the event horizon mapping into its viewed image. It is crucial that this image is a complete one-to-one projection of the whole event horizon sphere. This unique property of the event horizon image means that any distant observer may view at once the black hole on all sides. 

In this research we consider an idealized case for the event horizon imaging, when there are only compact luminous probes, plunging separately into black hole, and any other background or foreground luminous matter is absent. Meanwhile, we will use the position on the celestial sphere of the predicted black hole shadow in the Kerr metric for the corresponding localization of the event horizon image. In this respect it does not matter see you really the black hole shadow or not. 

In Fig.~4 the trajectories of some photons forming the event horizon image of the Schwarzschild black hole are shown, which are starting just above the event horizon globe: 4 (red colored) trajectories with $\sqrt{\lambda^2+q^2}=r_{\rm eh}$ are starting from the farthest point, and 4 (magenta colored) trajectories with $\sqrt{\lambda^2+q^2}=r_{\rm EW}$ are starting from the East-West meridian. Images of some parallels and meridians of the event horizon globe are also shown. A distant observer views at once both the front and back sides of the event horizon.

The outer boundary of event horizon image, viewed by a distant observer in the equatorial plane, is defined by solving the integral equation
\begin{equation}
\int_2^\infty\frac{dr}{\sqrt{V_r}}
=2\int_{\theta_0}^{\pi/2}\frac{d\theta}{\sqrt{V_\theta}}.
\label{a0max}
\end{equation}
where $\theta_0$ is a turning point in the latitudinal $\theta$-direction on the photon trajectory for direct image, defined from equation $V_\theta=0$. Photons, producing  the direct image in the Cunnungham--Bardeen classification scheme for multiple lensed images \cite{CunnBardeen72,CunnBardeen73}, do not intersect the black hole equatorial plane on the way from emitting probe to a distant observer.

In the simplest spherically symmetric Schwarzschild case (a=0) the event horizon sphere has radius $r_h=2$, a turning point $\theta_0=\arccos(q/\sqrt{q^2+\lambda^2})$. A corresponding value of the right-hand-side integral in (\ref{a0max}) is $\pi//\sqrt{q^2+\lambda^2}$. In result, from numerical solution of integral equation (\ref{a0max}) we find the radius of the event horizon image $r_{\rm eh}=\sqrt{q^2+\lambda^2}=4.457$. The near hemisphere of the event horizon with a radius $r_h=2$ is projected by the lensing photons into the central (light blue) disk of the image with a radius $r_{\rm EW}\simeq2.848$. Respectively the far hemisphere is projected into the hollow (dark blue) disk with an outer radius $r_{\rm eh}\simeq4.457$, which is a radius of the event horizon image. 

In the general axially symmetric Kerr case, when $a\neq0$, a turning point in $\theta$--direction is
\begin{equation}
\label{thetamax}
\theta_{0}\!=\!\arccos\!\left[\!\sqrt{\frac{\sqrt{4a^2q^2\!
			+\!(q^2\!+\!\lambda^2\!-\!a^2)^2}\!-\!(q^2\!
		+\!\lambda^2\!-\!a^2)}{2a^2}}\right]\!\!, \nonumber
\end{equation} 
and the event horizon image has a more complicated form. The corresponding numerical solution of integral equation  (\ref{a0max}) for the outer boundary of the event horizon image of the extreme Kerr black hole ($a=1$) is presented graphically in Figs.~6 and 7. The near hemisphere of the event horizon is projected by the lensing photons into the central (light blue) region.  Respectively the far hemisphere is projected into the hollow (dark blue) region. Images of some parallels and meridians of the event horizon globe are also shown.

The black hole horizon and its image are both synchronously rotating as a solid body with an angular velocity 
\begin{equation}
\Omega_{\rm h}=\frac{a}{2(1+\sqrt{1-a^2})}.
\label{OmegaH}
\end{equation}
Additionally the plunging test probes are infinitely winding in $\phi$ direction by approaching the event horizon if $a\neq0$. This very specific feature of the rotating black hole complicates the procedure for fixation of meridians on the event horizon image. By definition, the lensed image of meridians on the rotating event horizon globe are marked by photons, emitting at the same radii along the chosen meridian, $r=r_h+\epsilon$ and $\epsilon=const\ll1$, i.\,e., very close but still a little bit above the event horizon at $r=r_h$. Once being marked, these lensed image of meridian will be rotating synchronously with the event horizon. 

See in Fig.~8 the lensed images of compact star plunging into rotating black hole shown in discrete time (from numerical animation \cite{doknaz18c}). These lensed images are winding around black hole at the equatorial parallel $\theta=\pi/2$ of the event horizon globe when plunging star is approaching the event horizon. In particular, the last detected highly red-shifted photons from these lensed images reveal the equatorial parallel of the event horizon image which is placed inside the black hole shadow. For comparison see in Fig.~9 the numerical calculation of the direct image and the first and second light echoes images of a compact probe (star) orbiting around rotating black hole and viewed by a distant observer. All light echoes images are placed outside the black hole shadow.

\begin{figure}%[tbp]
	\centering
	\includegraphics[width=.45\textwidth,origin=c,angle=0]{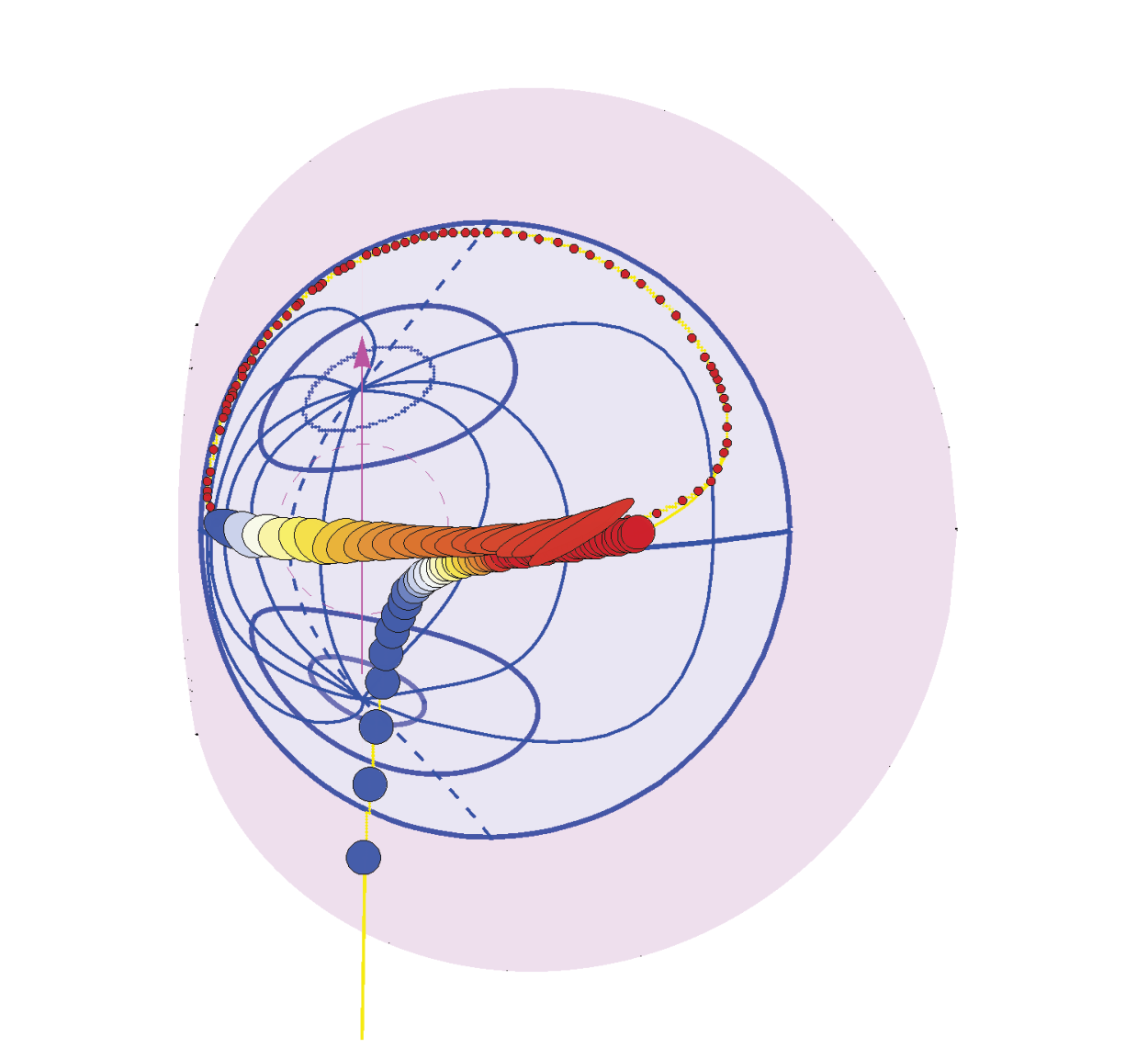}
	\caption{\label{fig:8} Numerical calculation of the lensed images of the spherical compact star with orbital parameters $\gamma=1$, $\lambda=q=0$ plunging into rotating black hole with $a=0.998$ at the equatorial plane and shown in discrete times. A distant observer is placed above the equatorial plane at $\cos\theta=0.1$. The lensed images are winding around black hole at the equatorial parallel $\theta=\pi/2$ of the event horizon globe when plunging star is approaching the event horizon. It is shown the first circle of this winding. See numerical animation of this gravitational lensing at \cite{doknaz18c}.}
\end{figure}
\begin{figure}%[tbp]
	\centering % \begin{center}/\end{center} takes some additional vertical space
	\includegraphics[width=0.45\textwidth,origin=c,angle=0]{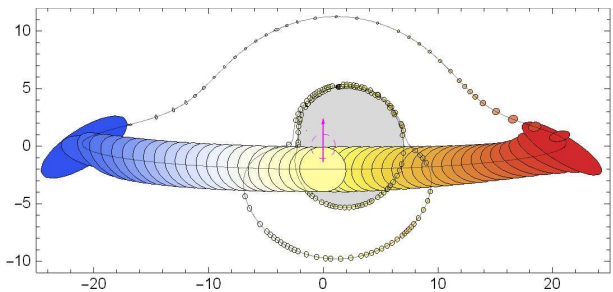}
	\caption{\label{fig:2} Gravitational lensing of a compact probe (star) orbiting the rotating black in the equatorial plane at radius $r=20$ and viewed by a distant observer in discrete time intervals. A distant observer is situated above the equatorial plane at $\cos\theta=0.1$. Direct star images and also the first and second light echoes around the black hole shadow (filled gray region) are shown in discrete time intervals.  For details see \cite{doknaz18a,doknaz18b}.}
\end{figure}

\section{Conclusion}

We show that gravitational lensing of the luminous matter plunging into black hole provides the principle possibility for visualization of the black hole event horizon. For a distant observer the event horizon image is projected on the celestial sphere within the corresponding black hole shadow. 

The event horizon image is projected within the position of black hole shadow on the celestial sphere, predicted by the Kerr metric (see Figs.~2--5). Depending on real astrophysical conditions, the future advanced observations of the supermassive black hole SgrA* have a chance to reveal both the black hole shadow and the event horizon image (by simultaneous monitoring with several telescopes of different types) or one of them. 

The black hole shadow is revealed by observations (e.\,g., in the X-rays or in the near far Infra-Red) of the foreground sources, e.\,g., the gravitationally lensed images of normal stars or neutron stars behind the black hole, which are moving around black hole on the stationary orbits. This possibility is illustrated in figure~2 and in the corresponding animated numerical simulation \cite{doknaz18a,doknaz18b}. 

At the same time, the distant observer may reveal event horizon image by detecting the highly red-shifted photons (e.\,g., in the radio waves) radiated very near the black hole horizon by the plunging compact sources (e.\,g., neutron stars or gas clouds). The helpful hints for the future observers of the black hole event horizon are the following: (1) Mark the position of the predicted black hole shadow on the celestial sphere by using Eqs.~(\ref{shadow1}) and (\ref{shadow2}) for the Kerr metric. It does not matter see you or not of this shadow in reality. (2) Wait for serendipitous appearance of the compact source plunging into black hole and provide monitoring of its lensed image within the marked black hole shadow. (3) The determined positions on the celestial sphere of the last detected highly red-shifted photons from different plunging sources reveal finally the requested event horizon image. This possibility is illustrated in figure~5 and in the corresponding numerical animation \cite{doknaz18c} of the lensed images of the spherical compact star plunging into rotating black hole at the equatorial plane. 

It must be stressed, however, that the described event horizon image may be recovered only under the very specific astrophysical conditions, when near the black hole there are only the rarely plunging compact objects and the stationary accretion flow is extremely small. The very promising case is the supermassive black hole SgrA* in the Galactic Center, which is a ``dormant quasar'' in view of its very weak nowadays activity.

A real visualization of the event horizon image is an extremely complicated problem because photons, emitted near the event horizon, are highly red-shifted by reaching the distant observer. For this reason the resulting accuracy of the event horizon imaging depends on the sensitivity of photon detector used by the distant observer. New technological solutions are requested for mining the highly red-shifted photons deeply buried inside the black hole shadow. 

The resulting event horizon image is a gravitationally lensed projection on the celestial sphere of the whole black hole event horizon sphere. Seemingly, black holes (as well as white holes and wormholes) are the unique objects in the Universe which may be viewed by distant observers at once from both the front and back sides. The similar 
statements on the properties of the event horizon image were made in \cite{Schwar} for the Schwarzschild case.

We finally conclude that a genuine image of the black hole is not its shadow, but a more compact event horizon image for observation of which the special conditions should be satisfied.

%\acknowledgments

We are grateful to E.~O.~Babichev, V.~A.~Berezin, Yu.~N.~Eroshenko and A.~L.~Smirnov for stimulating discussions. We thank the anonymous referee for constructive remarks which help to improve the presentation of the paper.

This work was supported in part by grant No.  18-52-15001 from Russian Foundation for Basic Research (RFBR) and Centre National de la Recherche Scientifique (CNRS).

\end{document}